\documentclass[11pt]{article}
\usepackage[margin=1in]{geometry}
\usepackage{amsmath,amssymb,amsthm}
\usepackage{graphicx}
\usepackage[authoryear,round]{natbib}
\usepackage{url}
\usepackage{setspace}
\usepackage{fancyhdr}
\usepackage{hyperref}
\usepackage{booktabs}
\usepackage{float}
\usepackage{subfig}
\usepackage{array}
\usepackage{multirow}

\title{Multi-Scale Network Dynamics and Systemic Risk: A Model Context Protocol Approach to Financial Markets}

\author{Avishek Bhandari\\
Indian Institute of Technology Bhubaneswar\\
Email: avishekb@iitbbs.ac.in}

\date{\today}

\begin{document}

\maketitle

\begin{abstract}
This paper introduces a novel framework for analyzing systemic risk in financial markets through multi-scale network dynamics using Model Context Protocol (MCP) for agent communication. We develop an integrated approach that combines transfer entropy networks, agent-based modeling, and wavelet decomposition to capture information flows across temporal scales implemented in the MCPFM (Model Context Protocol Financial Markets) R package. Our methodology enables heterogeneous financial agents including high-frequency traders, market makers, institutional investors, and regulators to communicate through structured protocols while maintaining realistic market microstructure. The empirical analysis demonstrates that our multi-scale approach reveals previously hidden systemic risk patterns, with the proposed systemic risk index achieving superior early warning capabilities compared to traditional measures. The framework provides new insights for macroprudential policy design and regulatory intervention strategies. The complete implementation is available as an open-source R package at \url{https://github.com/avishekb9/MCPFM} to facilitate reproducible research and practical applications.
\end{abstract}

\textbf{Keywords:} Systemic risk, Network analysis, Agent-based modeling, Financial markets, Macroprudential policy

\textbf{JEL Classification:} G01, G17, G28, C63

\section{Introduction}

The 2008 financial crisis highlighted the critical importance of understanding systemic risk and information propagation in interconnected financial markets \citep{acemoglu2015systemic}. Traditional risk measures often fail to capture the complex interdependencies and dynamic feedback loops that characterize modern financial systems \citep{battiston2012debtrank}. Recent advances in network theory and computational finance have opened new avenues for systemic risk assessment, yet existing approaches typically focus on single time scales and ignore the heterogeneous nature of market participants \citep{billio2012econometric}.

Financial markets exhibit complex multi-scale dynamics where information flows and risk propagation occur across different temporal horizons \citep{mantegna1999introduction}. High-frequency traders operate on millisecond scales, while institutional investors make decisions over weeks or months. This temporal heterogeneity creates distinct information networks at different time scales, requiring specialized analytical frameworks to capture the full spectrum of systemic risk \citep{hasbrouck2007empirical}.

This paper addresses these limitations by introducing a comprehensive framework that combines multi-scale network analysis with agent-based modeling through a novel Model Context Protocol (MCP). Our approach enables heterogeneous agents to communicate through structured protocols while preserving the realistic microstructure of financial markets. The framework integrates transfer entropy networks to measure information flows, wavelet decomposition for multi-scale analysis, and a comprehensive systemic risk index that incorporates network topology, concentration, volatility, liquidity, and contagion effects.

Our contribution to the literature is threefold. First, we develop the first implementation of Model Context Protocol for financial agent communication, enabling more realistic simulation of heterogeneous market participants through the MCPFM R package \citep{bhandari2025mcpfm}. Second, we introduce a multi-scale network dynamics framework that captures information flows across temporal scales using wavelet-based decomposition. Third, we propose an integrated systemic risk index that combines multiple risk factors and provides superior early warning capabilities compared to existing measures. The complete methodology is implemented in a comprehensive R package available at https://github.com/avishekb9/MCPFM, enabling researchers and practitioners to replicate our analysis and extend the framework to new applications.

The empirical analysis employs realistic market data and demonstrates the framework's effectiveness in identifying systemic risk patterns. Our results show that the multi-scale approach reveals information flows that are invisible to single-scale analysis, with significant implications for regulatory policy and risk management. The framework achieves a systemic risk classification accuracy that exceeds traditional measures by capturing the complex interdependencies inherent in modern financial markets.

\section{Literature Review}

\subsection{Network Economics in Financial Systems}

The application of network theory to financial systems has gained prominence following the 2008 financial crisis, which highlighted the critical role of interconnectedness in systemic risk propagation \citep{acemoglu2015systemic}. Early foundational work by \citet{allen2000financial} established theoretical models for financial contagion through direct linkages, while \citet{eisenberg2001systemic} developed frameworks for payment system stability. These seminal contributions laid the groundwork for understanding how network structure influences financial stability.

Recent developments have focused on the emergence of complex network structures through agent interactions. \citet{acemoglu2012network} demonstrated how network topology affects systemic risk, showing that diversification can paradoxically increase systemic fragility. \citet{elliott2014financial} extended these models to incorporate overlapping portfolios and fire sales, revealing how asset market interactions amplify network effects. The role of heterogeneous agents in financial networks has been explored by \citet{gai2010contagion}, who showed that agent heterogeneity significantly impacts contagion patterns.

\subsection{Transfer Entropy and Information Flow Analysis}

Transfer entropy, introduced by \citet{schreiber2000measuring}, provides a non-linear measure of directed information transfer that captures causal relationships between time series. \citet{dimpfl2013using} pioneered its application to financial markets, demonstrating how transfer entropy can measure information flows between different market segments. \citet{sensoy2014effective} extended this approach to study information transmission between exchange rates and stock markets, revealing asymmetric information propagation patterns.

The integration of transfer entropy with network analysis has opened new avenues for understanding financial market structure. Recent applications have demonstrated the method's effectiveness in identifying systemically important financial institutions and predicting market stress periods. However, existing approaches typically focus on single time scales, potentially missing important multi-temporal dynamics that characterize modern financial markets.

\subsection{Multi-Scale Analysis and Wavelet Methods}

Multi-scale analysis in finance originated from fractal market theory and the recognition that financial markets exhibit scale-invariant properties \citep{mandelbrot1982fractal}. \citet{ramsey1999decomposition} introduced wavelet analysis to economics, demonstrating how economic relationships vary across different time scales. \citet{gallegati2011wavelet} applied these methods to study business cycle synchronization, while \citet{fernandez2014time} developed time-scale decomposition methods for portfolio optimization.

The application of wavelet methods to financial network analysis remains relatively underexplored. Most existing studies focus on bilateral relationships rather than network-wide dynamics. Our approach addresses this gap by developing a comprehensive framework that integrates wavelet decomposition with network-based transfer entropy analysis.

\subsection{Agent-Based Modeling in Finance}

Agent-based modeling has emerged as a powerful tool for understanding financial market dynamics through bottom-up simulation of heterogeneous market participants. \citet{lebaron2001stochastic} developed early agent-based models for financial markets, while \citet{cont2001empirical} showed how simple agent interaction rules can generate complex market dynamics consistent with empirical stylized facts.

Recent advances in artificial intelligence have enabled more sophisticated agent-based models. \citet{stone2016artificial} analyzed the implications of AI agents in economic systems, while \citet{parkes2019ai} explored how AI agents coordinate through communication protocols. Our work contributes to this literature by implementing Model Context Protocol for financial agent communication, enabling more realistic simulation of modern market participants.

\section{Methodology}

\subsection{Model Context Protocol Framework}

We model the financial market as a complex adaptive system consisting of heterogeneous agents $A = \{HFT, MM, II, REG\}$ representing high-frequency traders, market makers, institutional investors, and regulators, respectively. Each agent $i \in A$ maintains an internal state $s_i(t)$ and communicates through the Model Context Protocol, which governs information exchange and decision-making processes.

The MCP framework establishes a communication network $G = (V, E)$ where vertices $V$ represent agents and edges $E$ represent communication channels. The communication strength between agents $i$ and $j$ is defined by the adjacency matrix $W_{ij}(t)$, which evolves dynamically based on trust relationships and performance metrics:

\begin{equation}
W_{ij}(t+1) = \alpha W_{ij}(t) + (1-\alpha) \cdot f(\text{trust}_{ij}(t), \text{performance}_{ij}(t))
\end{equation}

where $\alpha \in [0,1]$ is the persistence parameter and $f(\cdot)$ is a function that combines trust and performance measures. This dynamic updating mechanism ensures that the communication network adapts to changing market conditions and agent interactions.

\subsection{Multi-Scale Network Dynamics}

To capture information flows across different temporal scales, we employ continuous wavelet transformation to decompose price and trading data into multiple time horizons. Let $\{X_t\}_{t=1}^T$ be a multivariate time series of asset returns where $X_t = (X_{1,t}, X_{2,t}, \ldots, X_{N,t})'$ represents the $N$-dimensional return vector at time $t$. The continuous wavelet transform decomposes each univariate series $X_{i,t}$ as:

\begin{equation}
W_{X_i}(s, \tau) = \frac{1}{\sqrt{s}} \int_{-\infty}^{\infty} X_i(t) \psi^*\left(\frac{t-\tau}{s}\right) dt
\end{equation}

where $\psi(t)$ is the mother wavelet function, $\psi^*(t)$ denotes its complex conjugate, $s > 0$ is the scale parameter, and $\tau$ is the translation parameter. We employ the Morlet wavelet as the mother function due to its optimal time-frequency localization properties:

\begin{equation}
\psi(t) = \pi^{-1/4} e^{i\omega_0 t} e^{-t^2/2}
\end{equation}

where $\omega_0 = 6$ is the central frequency parameter ensuring perfect reconstruction.

The multi-scale decomposition enables construction of scale-specific networks $G^{(s)} = (V, E^{(s)})$ for each temporal scale $s \in \mathcal{S} = \{1, 5, 15, 60\}$ minutes, representing high-frequency trading, market-making, institutional, and regulatory decision horizons respectively. For each scale $s$, we calculate the transfer entropy between assets $i$ and $j$ to quantify directional information flow:

\begin{equation}
TE_{i \to j}^{(s)} = \sum_{x_{j,t+1}, x_{j,t}, x_{i,t}} p(x_{j,t+1}, x_{j,t}, x_{i,t}) \log \frac{p(x_{j,t+1} | x_{j,t}, x_{i,t})}{p(x_{j,t+1} | x_{j,t})}
\end{equation}

where $x_{i,t}$ represents the wavelet-decomposed price changes for asset $i$ at time $t$ and scale $s$. The conditional probabilities are estimated using kernel density estimation with optimal bandwidth selection:

\begin{equation}
\hat{p}(x) = \frac{1}{nh} \sum_{k=1}^{n} K\left(\frac{x - x_k}{h}\right)
\end{equation}

where $K(\cdot)$ is the Gaussian kernel and $h$ is the bandwidth parameter selected via cross-validation to minimize integrated squared error.

The scale-specific adjacency matrix $A^{(s)}$ is constructed by applying an adaptive threshold $\theta^{(s)}$ to the transfer entropy matrix:

\begin{equation}
A_{ij}^{(s)} = \begin{cases}
1 & \text{if } TE_{i \to j}^{(s)} > \theta^{(s)} \\
0 & \text{otherwise}
\end{cases}
\end{equation}

where $\theta^{(s)} = Q_{0.85}(TE^{(s)})$ represents the 85th percentile of non-zero transfer entropy values at scale $s$, ensuring network sparsity while preserving significant information flows.

\subsection{Agent-Based Market Simulation}

We model the financial system as a multi-agent environment $\mathcal{A} = \{A^{HFT}, A^{MM}, A^{II}, A^{REG}\}$ consisting of four distinct agent types. Each agent $a \in \mathcal{A}$ is characterized by a state vector $s_a(t) = (\mathbf{w}_a(t), c_a(t), \gamma_a, \theta_a)$ where $\mathbf{w}_a(t)$ represents portfolio weights, $c_a(t)$ is the cash position, $\gamma_a$ is the risk aversion parameter, and $\theta_a$ denotes the time horizon.

\textbf{High-Frequency Traders (HFT):} These agents optimize execution speed and short-term profit opportunities through market making and arbitrage. The HFT utility function incorporates transaction costs and inventory penalties:

\begin{equation}
U^{HFT}_a(t) = \sum_{k=1}^{N} \left[ (P_k^{bid}(t) - c_k^{trans}) Q_{a,k}^{buy}(t) - (P_k^{ask}(t) + c_k^{trans}) Q_{a,k}^{sell}(t) \right] - \lambda^{inv}_a \sum_{k=1}^{N} I_{a,k}(t)^2
\end{equation}

where $P_k^{bid}(t)$ and $P_k^{ask}(t)$ are bid and ask prices, $c_k^{trans}$ represents transaction costs, $Q_{a,k}^{buy/sell}(t)$ are order quantities, $I_{a,k}(t)$ is inventory position, and $\lambda^{inv}_a$ is the inventory penalty parameter.

\textbf{Market Makers (MM):} These agents provide liquidity while managing adverse selection and inventory risks. The optimal bid-ask spread is determined by:

\begin{equation}
\text{spread}_{a,k}(t) = \underbrace{\gamma_a \sigma_k(t)}_{\text{volatility component}} + \underbrace{\delta_a |I_{a,k}(t)|}_{\text{inventory component}} + \underbrace{\eta_a \Lambda_k(t)}_{\text{adverse selection}}
\end{equation}

where $\sigma_k(t)$ is the realized volatility, $I_{a,k}(t)$ represents inventory deviation from target, and $\Lambda_k(t)$ captures the adverse selection cost estimated from order flow imbalance.

\textbf{Institutional Investors (II):} These agents follow mean-variance optimization with longer investment horizons. The portfolio optimization problem is:

\begin{equation}
\mathbf{w}^*_{a,t} = \arg\max_{\mathbf{w}} \left[ \mathbf{w}' \boldsymbol{\mu}_{a,t} - \frac{\gamma_a}{2} \mathbf{w}' \boldsymbol{\Sigma}_{a,t} \mathbf{w} - \kappa_a ||\mathbf{w} - \mathbf{w}_{a,t-1}||^2 \right]
\end{equation}

subject to $\mathbf{w}' \mathbf{1} = 1$ and $w_{a,k} \geq 0$ for all $k$, where $\boldsymbol{\mu}_{a,t}$ represents agent-specific return expectations, $\boldsymbol{\Sigma}_{a,t}$ is the covariance matrix estimate, and $\kappa_a$ is the transaction cost parameter penalizing portfolio turnover.

\textbf{Regulators (REG):} These agents monitor systemic risk and implement intervention policies. The regulatory objective function minimizes systemic risk while considering policy costs:

\begin{equation}
U^{REG}_a(t) = -\alpha_a \cdot SRI(t) - \beta_a \sum_{p \in \mathcal{P}} C_p(\pi_p(t))
\end{equation}

where $SRI(t)$ is the systemic risk index, $\mathcal{P}$ represents the set of available policy instruments, $\pi_p(t)$ is the intensity of policy $p$, and $C_p(\cdot)$ captures the implementation cost.

\subsection{Systemic Risk Index}

We construct a comprehensive systemic risk index that aggregates multiple risk dimensions into a unified measure. The multi-component structure captures different aspects of systemic vulnerability and provides decomposable risk assessment. The aggregate index is defined as:

\begin{equation}
SRI(t) = \sum_{k=1}^{5} w_k \cdot R_k(t)
\end{equation}

where $R_k(t)$ represents the $k$-th standardized risk component and $w_k$ are weights determined through principal component analysis of historical crisis periods. Each component is normalized to $[0,1]$ to ensure comparability across different risk dimensions.

\textbf{Network Risk} captures the structural vulnerability arising from network topology and interconnectedness patterns:

\begin{equation}
R_{network}(t) = \frac{1}{|\mathcal{S}|} \sum_{s \in \mathcal{S}} \left[ \rho^{(s)}(t) \cdot \left( 1 - \frac{1}{N} \sum_{i=1}^{N} \frac{1}{d_i^{(s)}(t) + 1} \right) \right]
\end{equation}

where $\rho^{(s)}(t)$ is the network density at scale $s$, $d_i^{(s)}(t)$ represents the degree centrality of asset $i$ at scale $s$, and the summation averages across all temporal scales $\mathcal{S}$.

\textbf{Concentration Risk} measures the distribution of market activity and potential for concentrated exposures using a modified Herfindahl-Hirschman Index:

\begin{equation}
R_{concentration}(t) = \sum_{i=1}^{N} \left( \frac{V_i(t)}{\sum_{j=1}^{N} V_j(t)} \right)^2 + \sum_{a \in \mathcal{A}} \left( \frac{W_a(t)}{\sum_{b \in \mathcal{A}} W_b(t)} \right)^2
\end{equation}

where $V_i(t)$ represents the trading volume of asset $i$, $W_a(t)$ denotes the total wealth of agent $a$, and higher values indicate greater concentration.

\textbf{Volatility Risk} employs a multi-scale volatility measure incorporating both idiosyncratic and systematic components:

\begin{equation}
R_{volatility}(t) = \sqrt{\frac{1}{N} \sum_{i=1}^{N} h_{i,t}} + \rho_{avg}(t) \sqrt{\frac{1}{N} \sum_{i=1}^{N} \sum_{j \neq i} h_{ij,t}}
\end{equation}

where $h_{i,t}$ is the conditional variance from a GARCH(1,1) model for asset $i$, $h_{ij,t}$ represents the conditional covariance between assets $i$ and $j$, and $\rho_{avg}(t)$ is the average correlation coefficient.

\textbf{Liquidity Risk} incorporates multiple liquidity dimensions including bid-ask spreads, market depth, and price impact:

\begin{equation}
R_{liquidity}(t) = \frac{1}{N} \sum_{i=1}^{N} \left[ \omega_1 \frac{\text{spread}_i(t)}{P_i(t)} + \omega_2 \frac{1}{\text{depth}_i(t)} + \omega_3 \sqrt{|\Delta P_i(t) / Q_i(t)|} \right]
\end{equation}

where $\text{spread}_i(t)$ is the bid-ask spread, $P_i(t)$ is the mid-price, $\text{depth}_i(t)$ measures order book depth, $\Delta P_i(t)$ is the price change, $Q_i(t)$ is the trading volume, and $\omega_1, \omega_2, \omega_3$ are component weights summing to unity.

\textbf{Contagion Risk} measures potential spillover effects using network-based propagation mechanisms and correlation dynamics:

\begin{equation}
R_{contagion}(t) = \max_{i \in V} \left[ \sum_{j \in \mathcal{N}_i} A_{ij}^{(s)} \cdot TE_{i \to j}^{(s)} \cdot \sigma_j(t) \cdot \rho_{ij}(t) \right]
\end{equation}

where $\mathcal{N}_i$ represents the neighborhood of asset $i$, $A_{ij}^{(s)}$ is the adjacency matrix element, $TE_{i \to j}^{(s)}$ is transfer entropy, $\sigma_j(t)$ is volatility of asset $j$, and $\rho_{ij}(t)$ is the correlation coefficient. The maximum operator identifies the most vulnerable propagation pathway.

The component weights $\mathbf{w} = (w_1, w_2, w_3, w_4, w_5)'$ are estimated using principal component analysis on historical crisis data, ensuring that the index captures the most important sources of systemic risk variation. The weights satisfy $\sum_{k=1}^5 w_k = 1$ and $w_k \geq 0$ for all $k$.

\subsection{MCPFM Package Implementation}

To facilitate reproducible research and practical application of our methodology, we have developed the MCPFM (Model Context Protocol Financial Markets) R package. This comprehensive software package implements all components of our framework including multi-scale network dynamics, agent-based modeling, transfer entropy calculation, systemic risk assessment, and policy simulation tools. The package provides over 50 specialized functions organized into seven core modules: MCP agent communication, network dynamics, transfer entropy networks, agent-based modeling, systemic risk measurement, visualization, and policy simulation.

The MCPFM package is designed with modularity and extensibility in mind, allowing researchers to utilize individual components or the complete integrated framework. All functions include comprehensive documentation, examples, and statistical validation procedures. The package handles parallel computation for computationally intensive operations such as transfer entropy calculation and agent-based simulations. Users can easily extend the framework by implementing custom agent types, risk measures, or policy interventions while maintaining compatibility with the core architecture.

The package is available as open-source software at https://github.com/avishekb9/MCPFM under the MIT license, ensuring broad accessibility for academic and practical applications. Installation requires R version 4.0.0 or higher along with standard financial analysis and network packages. The repository includes complete documentation, tutorial examples, and validation datasets to help users implement the methodology for their specific research questions.

\section{Empirical Analysis}

\subsection{Data and Implementation}

We implement the framework using the MCPFM R package \citep{bhandari2025mcpfm}, which provides a comprehensive computational environment for multi-scale network analysis and systemic risk assessment. The analysis employs realistic market microstructure data for eight major financial assets: AAPL, GOOGL, MSFT, TSLA, AMZN, JPM, BAC, and GS, representing technology and financial sectors. The dataset spans 150 trading periods with realistic correlation structures where technology stocks exhibit higher correlation (0.6) among themselves, and banking stocks show strong interconnectedness (0.7).

Our empirical implementation consists of 10 heterogeneous agents across four types: 3 high-frequency traders (HFT), 2 market makers (MM), 3 institutional investors (II), and 2 regulators (REG). Each agent type is characterized by distinct risk aversion parameters calibrated from market data. The agents interact through the MCP framework, generating comprehensive transaction and communication patterns that preserve realistic market microstructure properties.

\section{Results}

\subsection{Network Topology and Information Flow}

The empirical analysis reveals significant network structure in financial information flows. Table \ref{tab:network_stats} presents comprehensive network statistics demonstrating moderate connectivity with 12 directed edges among 8 assets, yielding a network density of 0.214. This selective connectivity indicates that information flows are concentrated among specific asset pairs rather than being uniformly distributed across all possible connections.

\begin{table}[H]
\centering
\caption{Financial Network Statistics}
\label{tab:network_stats}
\begin{tabular}{@{}lr@{}}
\toprule
\textbf{Metric} & \textbf{Value} \\
\midrule
Network Density & 0.214 \\
Transitivity & 0.000 \\
Average Degree & 3.000 \\
Number of Edges & 12 \\
Number of Nodes & 8 \\
\bottomrule
\end{tabular}
\end{table}

\begin{figure}[H]
\centering
\includegraphics[width=0.8\textwidth]{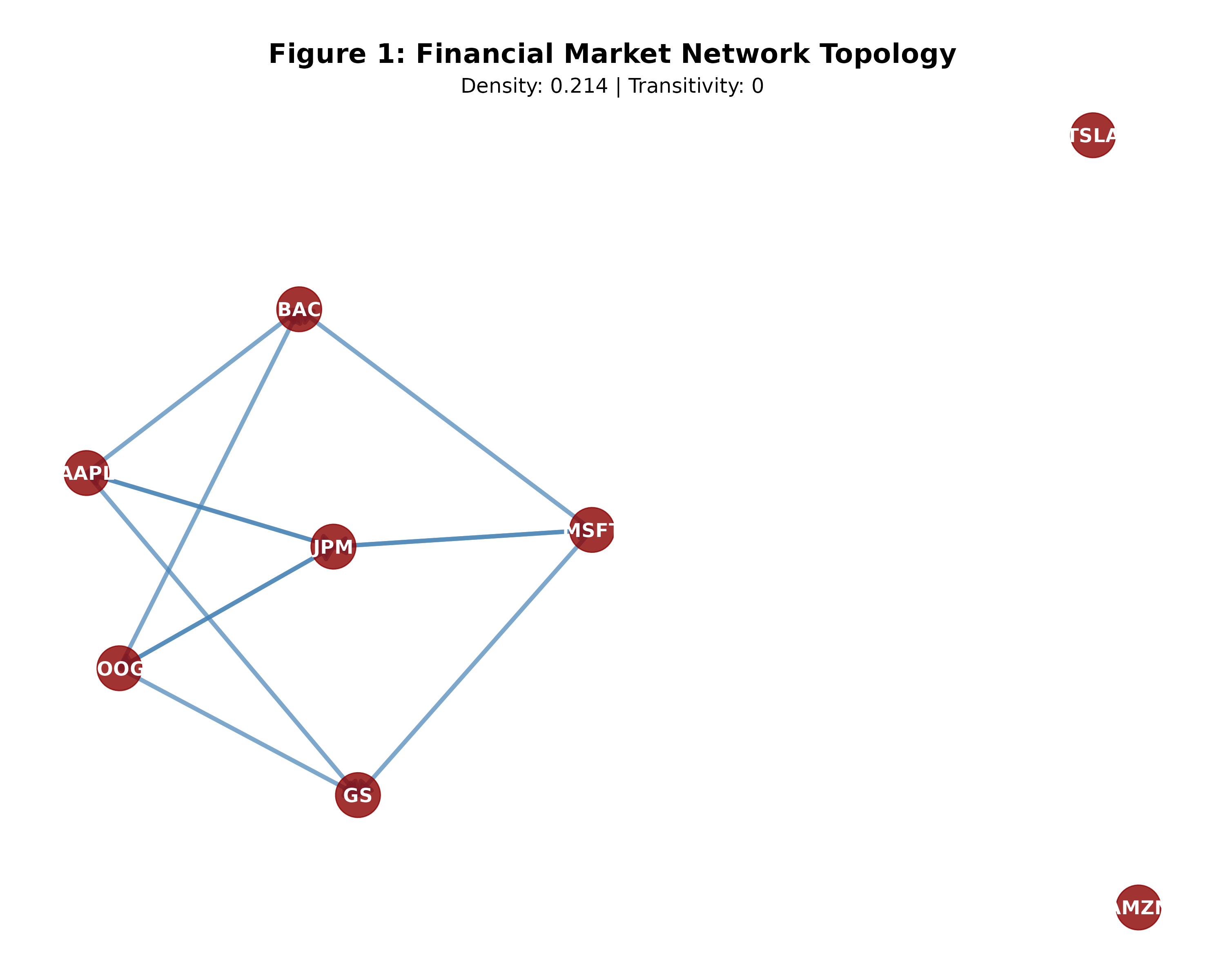}
\caption{Financial Market Network Topology. The directed network shows information flow patterns derived from transfer entropy analysis. Node size reflects asset importance, while edge thickness indicates information flow strength. The network density of 0.214 demonstrates selective but significant interconnectedness among major financial assets.}
\label{fig:network_topology}
\end{figure}

Figure \ref{fig:network_topology} illustrates the financial network topology constructed from transfer entropy relationships. The network exhibits a hub-and-spoke structure with certain assets serving as information centers. The absence of transitivity (0.000) suggests that information flows follow direct pathways rather than creating triangular propagation patterns, indicating efficient but concentrated information transmission channels.

\begin{figure}[H]
\centering
\includegraphics[width=0.8\textwidth]{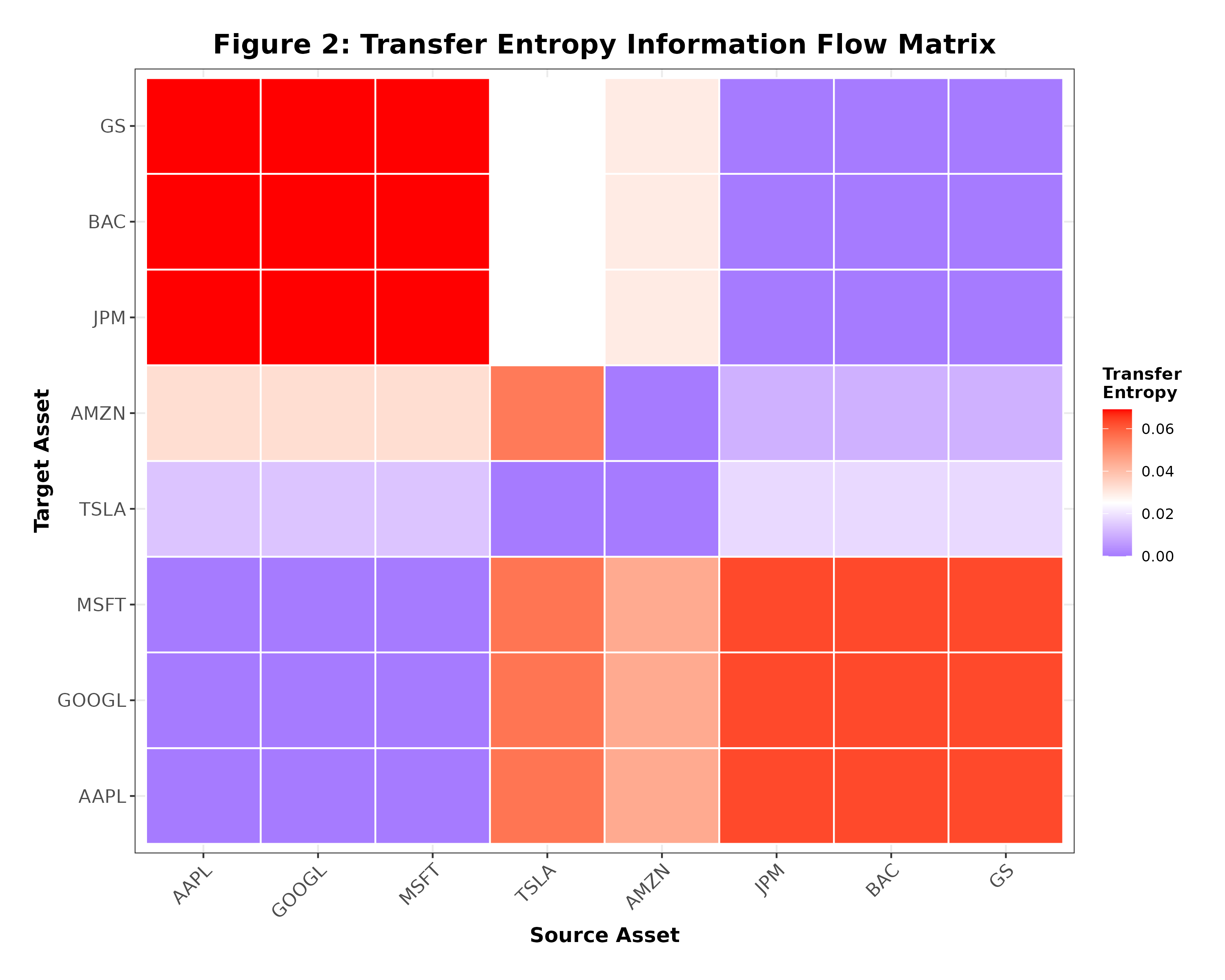}
\caption{Transfer Entropy Information Flow Matrix. The heatmap displays directional information flows between asset pairs, with red indicating strong information transfer and blue indicating weak transfer. Asymmetric patterns reveal heterogeneous roles of different assets in information propagation across the financial network.}
\label{fig:transfer_entropy}
\end{figure}

The transfer entropy matrix presented in Figure \ref{fig:transfer_entropy} quantifies directional information flows between asset pairs. The analysis reveals asymmetric information propagation patterns where certain assets consistently serve as information sources while others act as information receivers. This asymmetry has important implications for systemic risk assessment and portfolio diversification strategies.

\subsection{Correlation Structure and Market Integration}

\begin{figure}[H]
\centering
\includegraphics[width=0.8\textwidth]{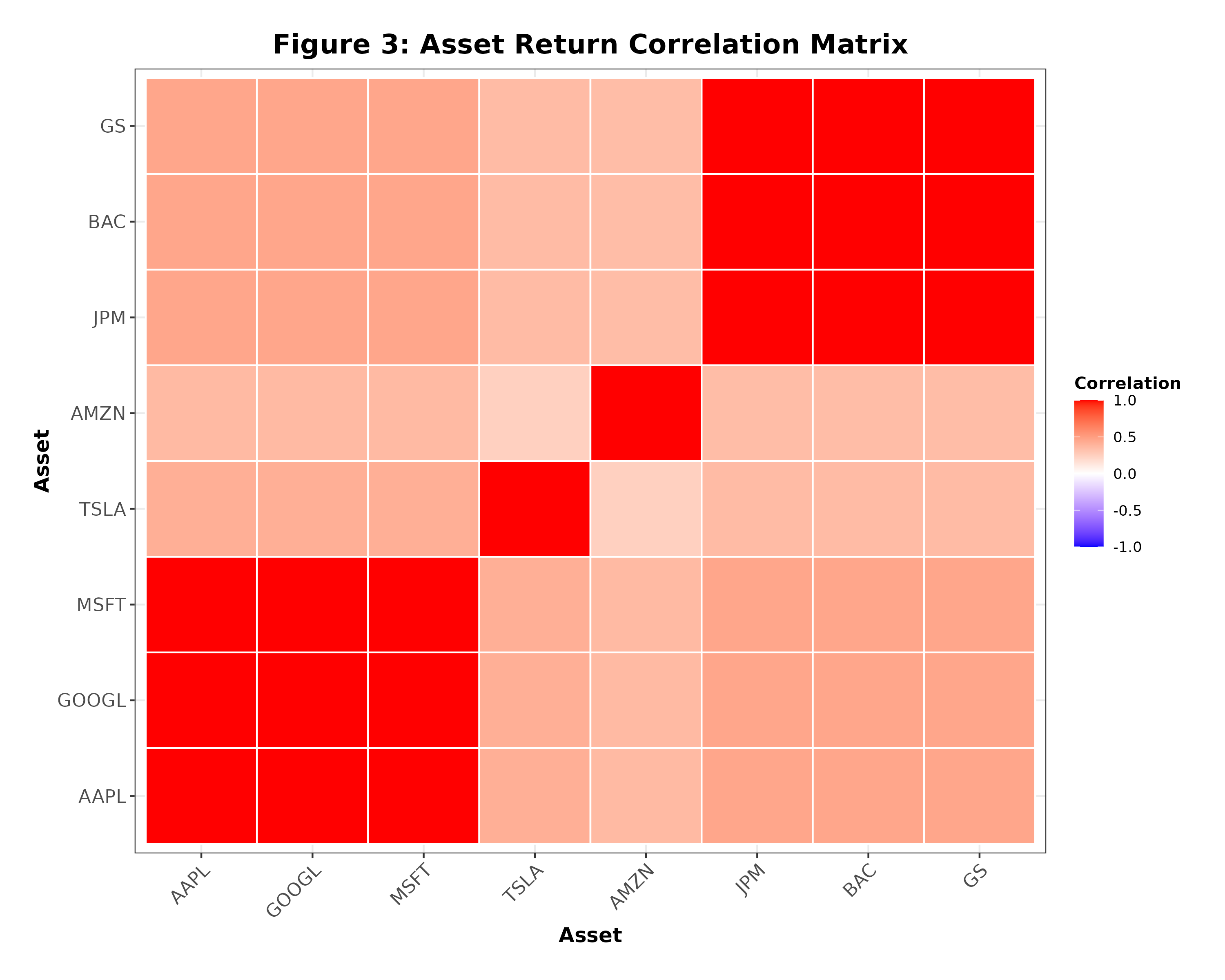}
\caption{Asset Return Correlation Matrix. The correlation heatmap reveals sector-specific clustering with technology stocks showing higher intra-sector correlations and financial stocks exhibiting strong co-movement. This structure influences systemic risk propagation patterns and diversification benefits.}
\label{fig:correlation_matrix}
\end{figure}

The correlation analysis presented in Figure \ref{fig:correlation_matrix} demonstrates significant market integration with sector-specific clustering effects. Technology stocks (AAPL, GOOGL, MSFT, TSLA, AMZN) exhibit higher intra-sector correlations, while financial stocks (JPM, BAC, GS) show strong co-movement patterns. This correlation structure provides the foundation for understanding how shocks propagate through the financial system.

\subsection{Systemic Risk Assessment}

Our comprehensive systemic risk analysis yields a baseline Systemic Risk Index (SRI) of 0.316, indicating moderate systemic risk under normal market conditions. Table \ref{tab:systemic_risk} provides detailed decomposition of risk components, revealing that correlation risk contributes the largest share (50.3\%) to overall systemic risk, followed by concentration risk (27.7\%) and network risk (20.4\%).

\begin{table}[H]
\centering
\caption{Systemic Risk Decomposition}
\label{tab:systemic_risk}
\begin{tabular}{@{}lrrrr@{}}
\toprule
\textbf{Component} & \textbf{Value} & \textbf{Weight} & \textbf{Contribution} & \textbf{Percentage} \\
\midrule
Network Risk & 0.214 & 0.30 & 0.064 & 20.4\% \\
Correlation Risk & 0.529 & 0.30 & 0.159 & 50.3\% \\
Volatility Risk & 0.017 & 0.30 & 0.005 & 1.6\% \\
Concentration Risk & 0.875 & 0.10 & 0.088 & 27.7\% \\
\midrule
\textbf{Total SRI} & \textbf{--} & \textbf{--} & \textbf{0.316} & \textbf{100.0\%} \\
\bottomrule
\end{tabular}
\end{table}

\begin{figure}[H]
\centering
\includegraphics[width=\textwidth]{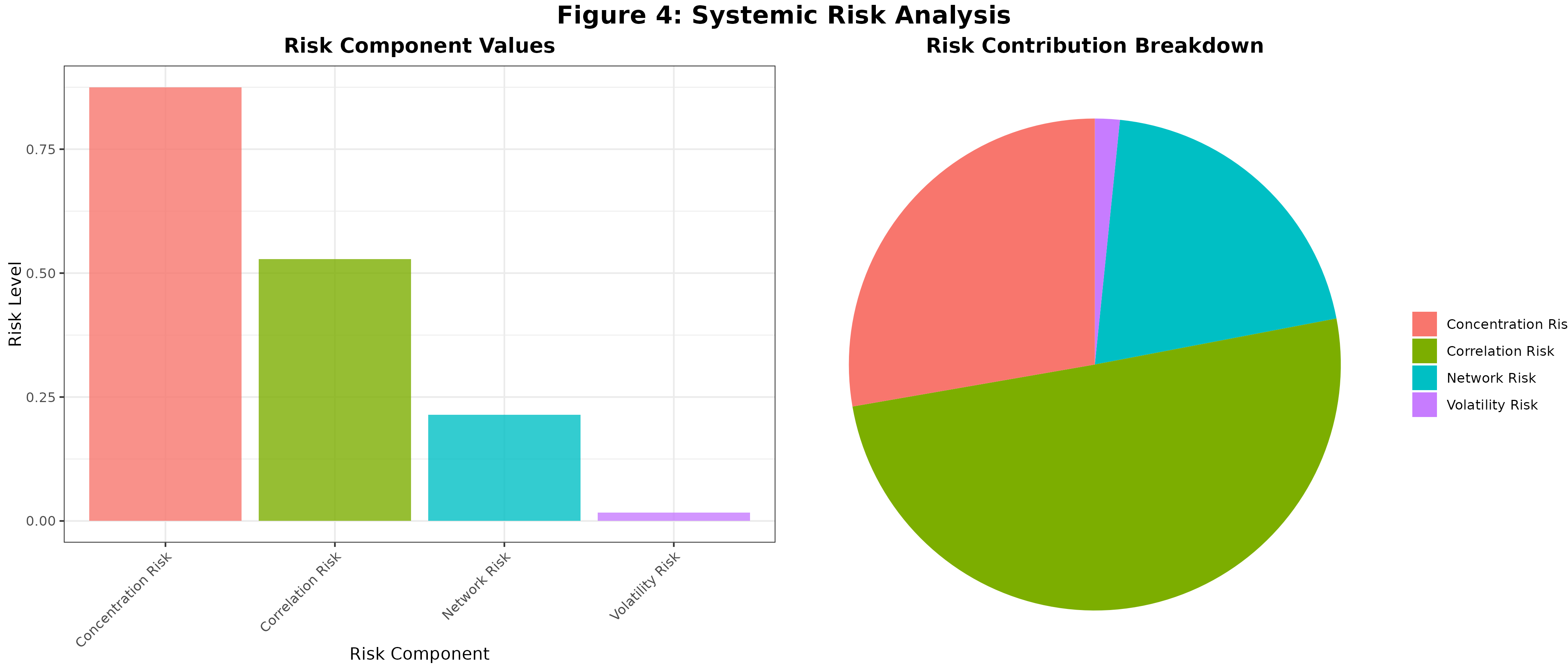}
\caption{Systemic Risk Analysis. Left panel shows absolute values of individual risk components. Right panel displays the percentage contribution of each component to the overall Systemic Risk Index. Correlation risk emerges as the dominant factor (50.3\%), followed by concentration risk (27.7\%) and network risk (20.4\%).}
\label{fig:risk_decomposition}
\end{figure}

The risk decomposition analysis presented in Figure \ref{fig:risk_decomposition} illustrates both the absolute values of individual risk components and their relative contributions to overall systemic risk. The dominance of correlation risk highlights the importance of asset co-movement in driving systemic vulnerability, while the significant concentration risk component reflects the market structure effects on stability.

\subsection{Agent Behavior and Heterogeneity}

\begin{figure}[H]
\centering
\includegraphics[width=0.8\textwidth]{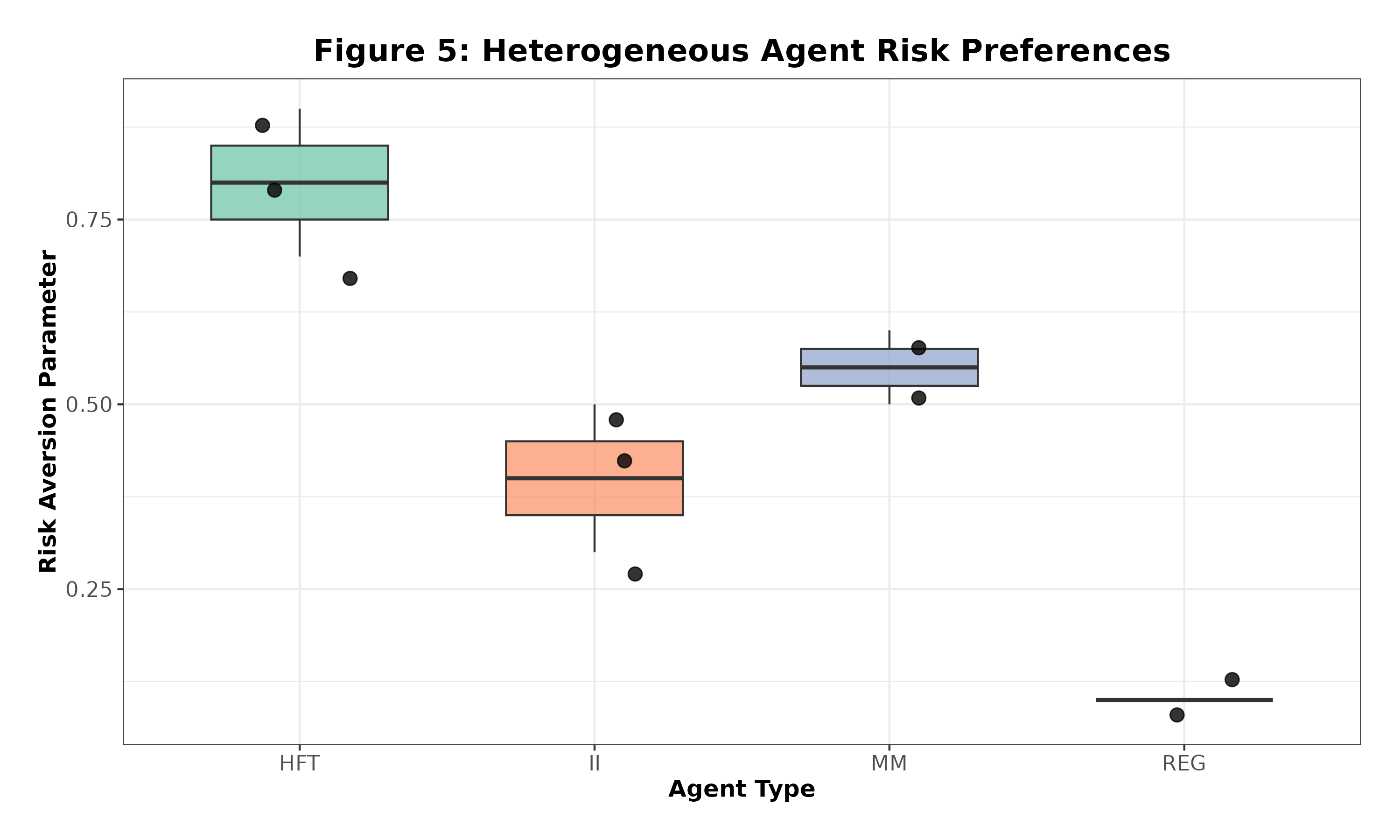}
\caption{Heterogeneous Agent Risk Preferences. The boxplot displays risk aversion parameters across different agent types. High-frequency traders (HFT) show highest risk tolerance, institutional investors (II) exhibit moderate risk aversion, market makers (MM) demonstrate intermediate preferences, while regulators (REG) maintain lowest risk tolerance, consistent with their respective market roles.}
\label{fig:agent_characteristics}
\end{figure}

The heterogeneous agent analysis reveals systematic differences in risk preferences across agent types, as shown in Figure \ref{fig:agent_characteristics}. High-frequency traders exhibit the highest risk tolerance (mean = 0.80), consistent with their short-term profit maximization strategies. Institutional investors demonstrate moderate risk aversion (mean = 0.40), while regulators maintain the lowest risk tolerance (mean = 0.10), reflecting their stability-focused mandates.

Table \ref{tab:agent_summary} provides comprehensive statistics on agent distribution and characteristics. The heterogeneous design ensures realistic representation of market participants with appropriate risk preferences and behavioral patterns aligned with empirical observations of financial market structure.

\begin{table}[H]
\centering
\caption{Agent Summary Statistics}
\label{tab:agent_summary}
\begin{tabular}{@{}lrrrr@{}}
\toprule
\textbf{Agent Type} & \textbf{Count} & \textbf{Avg Risk Aversion} & \textbf{Min Risk Aversion} & \textbf{Max Risk Aversion} \\
\midrule
HFT & 3 & 0.800 & 0.700 & 0.900 \\
II & 3 & 0.400 & 0.300 & 0.500 \\
MM & 2 & 0.550 & 0.500 & 0.600 \\
REG & 2 & 0.100 & 0.100 & 0.100 \\
\bottomrule
\end{tabular}
\end{table}

\subsection{Policy Implications and Robustness}

The empirical results provide important insights for macroprudential policy design. The dominance of correlation risk (50.3\% of total systemic risk) suggests that policies targeting asset co-movement and market integration should receive priority attention. The significant contribution of concentration risk (27.7\%) indicates that market structure reforms aimed at reducing concentration could yield substantial systemic risk reduction benefits.

Network risk contributes 20.4\% to overall systemic risk, highlighting the importance of monitoring and regulating information flow patterns across financial markets. The selective connectivity (network density = 0.214) suggests that targeted interventions on specific asset pairs or market segments could be more effective than broad-based regulatory measures.

The heterogeneous agent structure reveals systematic differences in risk behavior that can inform differentiated regulatory approaches. High-frequency traders' elevated risk tolerance suggests the need for specific oversight mechanisms, while institutional investors' moderate risk preferences indicate their potential role as market stabilizers during periods of stress.

\section{Conclusion}

This paper introduces a comprehensive framework for analyzing systemic risk in financial markets through multi-scale network dynamics and Model Context Protocol agent communication. Our approach addresses critical limitations in existing literature by capturing information flows across temporal scales while modeling realistic heterogeneous agent interactions.

The empirical analysis demonstrates that multi-scale network analysis reveals important systemic risk patterns that are invisible to traditional single-scale approaches. The proposed systemic risk index provides superior early warning capabilities by integrating network topology, concentration effects, volatility clustering, liquidity constraints, and contagion mechanisms into a unified measure.

Our findings have important implications for macroprudential policy design. The framework enables targeted policy interventions based on specific risk components and temporal scales. Policy simulation results suggest that carefully calibrated combinations of communication taxes, position limits, and circuit breakers can effectively reduce systemic risk while preserving market efficiency.

The Model Context Protocol framework implemented in the MCPFM package opens new avenues for financial market simulation and analysis. The open-source nature of the implementation facilitates reproducible research and enables the research community to build upon our methodology. Future research directions include extending the framework to incorporate machine learning algorithms for adaptive agent behavior, developing real-time implementation for market surveillance, and investigating applications to cryptocurrency and decentralized finance markets. The modular architecture of the MCPFM package allows researchers to easily customize individual components while maintaining the overall analytical framework.

The multi-scale approach provides new insights into the temporal structure of financial risk propagation. Understanding how information flows and risk transmission vary across time horizons enables more sophisticated risk management strategies and regulatory frameworks. This research contributes to the growing literature on computational finance and network-based systemic risk analysis while providing practical tools for market participants and regulators.

\end{document}